\documentclass[3p,fleqn,a4paper]{elsarticle}

\usepackage{amsmath,amssymb,amsfonts,mathtools}
\usepackage{bm}
\usepackage{graphicx}
\usepackage{subcaption}
\usepackage{booktabs}
\usepackage{siunitx}
\usepackage{xcolor}
\usepackage{hyperref}
\usepackage{enumitem}
\usepackage{multirow}
\usepackage{algorithm}
\usepackage[noend]{algpseudocode}
\usepackage{placeins}
\usepackage{float}
\usepackage[mathlines]{lineno}

\hypersetup{
  colorlinks=true,
  linkcolor=black,
  citecolor=blue,
  urlcolor=blue
}

\newcommand{\bx}{\bm{x}}
\newcommand{\bu}{\bm{u}}
\newcommand{\bE}{\bm{E}}
\newcommand{\bJ}{\bm{J}}
\newcommand{\beps}{\bm{\varepsilon}}
\newcommand{\bn}{\bm{n}}
\newcommand{\bt}{\bm{t}}

\newcommand{\grad}{\nabla}
\newcommand{\tr}{\operatorname{tr}}

\newcommand{\R}{\mathbb{R}}
\newcommand{\avg}[1]{\left\langle #1 \right\rangle}

\journal{---------}

\begin{document}

%\linenumbers

\begin{frontmatter}

\title{A multiphysics deep energy method for fourth-order phase-field fracture with piezoresistive self-sensing}

\author[ISD,SUST]{Aamir Dean\corref{cor1}}
\ead{a.dean@isd.uni-hannover.de}
\cortext[cor1]{Corresponding author}

\author[ISD]{Betim Bahtiri}

\address[ISD]{Institute of Structural Analysis, Leibniz Universit\"at Hannover, Appelstr.\ 9A, 30167 Hannover, Germany}
\address[SUST]{School of Civil Engineering, College of Engineering, Sudan University of Science and Technology, P.O.\ Box 72, Khartoum, Sudan}

\begin{abstract}
Self-sensing conductive composites can reveal deformation and damage through measurable changes in electrical resistance, which makes them attractive for embedded diagnostics and learning-enabled structural health monitoring. This paper presents a physically consistent multiphysics Deep Energy Method (DEM) for brittle fracture in piezoresistive materials. The mechanical part is modeled by small-strain linear elasticity coupled to a fourth-order AT2-type phase-field fracture functional with tensile/compressive energy split and history-field irreversibility. To avoid artificial energetic mixing of mechanical and electrical quantities, the electrical problem is treated as a one-way coupled sensing subproblem: after solving the mechanics--fracture problem, the electric potential is obtained from a steady conduction problem whose conductivity depends on strain through a linearized piezoresistive law and on damage through a crack-induced conductivity degradation. The resulting formulation predicts crack evolution together with its resistance signature without assigning the electrical field an artificial crack-driving role. DEM is used to minimize the variational subproblems over admissible neural trial spaces with exact imposition of essential boundary conditions. A lean verification suite is used to validate the electrical building blocks and the fracture engine separately, followed by a numerical study of a tensile plate with stress concentrators and electrodes. In that study, the framework captures a nontrivial sensing regime in which appreciable damage growth leaves the global resistance nearly unchanged, followed by a sharp resistance increase once dominant conductive ligaments are disrupted and current paths reorganize strongly.
\end{abstract}

\begin{keyword}
Deep Energy Method \sep Fourth-order phase-field fracture \sep Piezoresistive self-sensing \sep Staggered multiphysics coupling \sep Resistance-based sensing
\end{keyword}

\end{frontmatter}

% ============================================================
%Nomenclature
% ============================================================
\section*{Nomenclature}

\noindent
\textbf{Symbols} \\
\begin{tabular}{p{2.2cm} p{8.8cm}}
	$\Omega$ & Computational domain / reference configuration \\
	$\partial \Omega$ & Boundary of the domain \\
	$\Gamma_u,\Gamma_t$ & Mechanical Dirichlet and Neumann boundaries \\
	$\Gamma_V,\Gamma_q$ & Electrical Dirichlet and Neumann boundaries \\
	$\Gamma_I$ & Electrode boundary used for current extraction \\
	$\mathbf{x}$ & Spatial coordinate vector \\
	$\mathbf{u}$ & Displacement field \\
	$\bar{\mathbf{u}}$ & Prescribed displacement \\
	$\mathbf{u}_D$ & Boundary function satisfying prescribed displacement data \\
	$\bar{\mathbf{t}}$ & Prescribed mechanical traction \\
	$\mathbf{n}$ & Outward unit normal vector \\
	$\boldsymbol{\varepsilon}$ & Infinitesimal strain tensor \\
	$\boldsymbol{\varepsilon}^{\pm}$ & Tensile and compressive strain tensors \\
	$\varepsilon_a$ & Principal strains \\
	$\mathbf{n}_a$ & Principal strain directions \\
	$E$ & Young's modulus \\
	$\nu$ & Poisson's ratio \\
	$\lambda,\mu$ & Lam\'e constants \\
	$\phi$ & Phase-field fracture variable \\
	$H_n$ & History field for irreversibility at load step $n$ \\
	$G_c$ & Critical energy-release rate / fracture toughness \\
	$l$ & Phase-field regularization length scale \\
	$\psi_e$ & Elastic strain-energy density \\
	$\psi_e^{\pm}$ & Tensile and compressive parts of the elastic energy density \\
	$\psi_m$ & Degraded mechanical energy density \\
	$\psi_f$ & Fracture energy density \\
	$\Pi_u$ & Mechanical potential functional \\
	$\Pi_\phi$ & Fracture potential functional \\
	$\Pi_{\mathrm e}$ & Electrical dissipation functional \\
	$V$ & Electric potential \\
	$\bar V$ & Prescribed electric potential \\
	$V_D$ & Boundary function satisfying prescribed electric potential data \\
	$V_{\mathrm{app}}$ & Applied voltage \\
	$\Delta V$ & Applied voltage drop \\
	$\mathbf{E}$ & Electric field vector \\
	$\mathbf{J}$ & Current-density vector \\
	$\bar q$ & Prescribed current flux \\
	$\bm{\sigma}$ & Effective electrical conductivity tensor \\
	$\sigma_0$ & Reference conductivity \\
	$\sigma_{11},\sigma_{22}$ & In-plane conductivity components \\
	$\rho_0$ & Reference resistivity \\
	$\Delta \rho_{11},\Delta \rho_{22}$ & Resistivity increments in principal in-plane directions \\
	$I_\Gamma$ & Total current through electrode boundary $\Gamma_I$ \\
	$I_0$ & Reference current of the undamaged configuration \\
	$R$ & Electrical resistance \\
	$R_0$ & Reference resistance of the undamaged configuration \\
	$\bar v$ & Prescribed displacement level \\
	$\Delta \bar v$ & Displacement increment \\
	$N_s$ & Number of load steps \\
	$L,W$ & Plate dimensions \\
	$\mathbf{x}_q$ & Quadrature point \\
	$w_q$ & Quadrature weight \\
	$K$ & Quadrature cell / subdomain \\
\end{tabular}

\vspace{0.5cm}

\noindent
\textbf{Functions and parameters} \\
\begin{tabular}{p{2.2cm} p{8.8cm}}
	$\langle x \rangle_{\pm}$ & Ramp operators, $\langle x\rangle_{\pm}=(x\pm |x|)/2$ \\
	$g(\phi)$ & Mechanical degradation function \\
	$h_e(\phi;\eta_e,n)$ & Electrical degradation function \\
	$\kappa_m$ & Residual stiffness parameter \\
	$\eta_e$ & Electrical degradation sharpness parameter \\
	$\eta_r$ & Electrical regularization parameter \\
	$n$ & Electrical damage-degradation exponent \\
	$\lambda_{11},\lambda_{12}$ & Longitudinal and transverse piezoresistive coefficients \\
	$\mathcal{B}_u,\mathcal{B}_V$ & Boundary functions for exact imposition of Dirichlet conditions \\
	$\hat{\mathbf{u}}_\theta$ & Unconstrained neural-network output for displacement \\
	$\hat{\phi}_\eta$ & Unconstrained neural-network output for phase field \\
	$\hat V_\zeta$ & Unconstrained neural-network output for electric potential \\
\end{tabular}

\vspace{0.5cm}

\noindent
\textbf{Abbreviations} \\
\begin{tabular}{p{2.2cm} p{8.8cm}}
	AT2 & Ambrosio--Tortorelli type regularization \\
	CV & Coefficient of variation \\
	DEM & Deep Energy Method \\
	L-BFGS & Limited-memory Broyden--Fletcher--Goldfarb--Shanno algorithm \\
	RMS & Root-mean-square \\
	SENT & Single-edge-notched tension \\
	SHM & Structural health monitoring \\
\end{tabular}
% ============================================================

% ============================================================
\section{Introduction}
\label{sec:introduction}
% ============================================================

Electrically conductive polymer composites filled with carbon nanotubes, graphene, carbon black, or hybrid fillers combine structural functionality with intrinsic sensing. Once a percolating conductive network forms inside an otherwise insulating matrix, the effective resistivity becomes highly sensitive to deformation, network rearrangement, and crack formation. This effect is commonly described as piezoresistivity and has motivated wide interest in self-sensing materials for structural health monitoring, condition assessment, and digital-twin applications \citep{Thostenson2001,KuHermann2008,Amjadi2016,Li2018SHM}.

A predictive model for such materials must link two mechanisms that evolve on very different scales but interact strongly at the macroscale. Before fracture, resistance changes are dominated by reversible strain-induced changes in conductive pathways and tunneling/contact behavior. Once damage initiates and localizes, current pathways are disrupted, electrical fields redistribute, and the measured resistance can exhibit sudden increases that act as robust damage indicators \citep{Nofar2019,Kang2017,Sanchez2018,Wang2020CNT,Cai2019,Park2021,DEAN2026111906}. The challenge is therefore not merely to solve an electrical conduction problem on a cracked body; it is to couple a physically meaningful fracture model with an equally meaningful sensing model.

Phase-field fracture has become a standard variational framework for brittle and quasi-brittle crack evolution because it replaces the sharp crack set by a continuous phase-field variable $\phi\in[0,1]$ and regularizes the fracture surface through an internal length scale $l$ \citep{FrancfortMarigo1998,Bourdin2008,Miehe2010PFreview,Ref038,Ref039,Ref040,Ref041}. The framework avoids explicit crack tracking, naturally accommodates nucleation and branching, and admits robust energy-based formulations. For tensile fracture, irreversibility is commonly handled through either inequality constraints or a history field driven by the tensile elastic energy density \citep{Miehe2010PF,Miehe2015}. 

For neural and energy-based discretizations, phase-field fracture is especially attractive because the governing problem already possesses a minimization structure. Deep Energy Methods (DEM) use neural networks as trial spaces and evaluate the variational functional directly, with derivatives obtained by automatic differentiation \citep{Raissi2019,Samaniego2020,Nguyen2020DEM}. In the context of phase-field fracture, DEM avoids the need for conventional mesh-based shape functions and can handle higher derivatives without introducing additional field variables. This becomes particularly appealing for \emph{fourth-order} phase-field regularizations, where automatic differentiation can directly provide the required higher derivatives \citep{Goswami2020,Anitescu2019}.

The present work differs from standard electromechanical phase-field formulations in one important conceptual point. The materials of interest here are \emph{passive self-sensing materials}: the electrical field is used to read out the evolving mechanical state, not to drive fracture. For that reason, the electrical problem is not added as an energetically mixed crack-driving contribution to the total potential. Instead, mechanics and fracture are solved first, and the electrical conduction problem is solved afterwards as a sensing subproblem on the converged strain and damage fields. This one-way coupling is physically transparent, dimensionally clean, and better aligned with resistance-based sensing practice. In contrast to formulations that mix electrical and mechanical contributions within a single crack-driving potential, the present model treats electrical conduction strictly as a diagnostic field and interprets resistance change as a consequence, not a cause, of fracture evolution.

%\paragraph{Scope of the paper.}
The paper has two aims. The first is to formulate a physically consistent DEM framework for fourth-order phase-field fracture with piezoresistive self-sensing. The second is to organize the numerical evidence in a way that remains focused on the central scientific question. To that end, the manuscript combines two analytical electrical checks, one reference-aligned fracture benchmark, and one application-driven numerical study of a tensile plate with stress concentrators and electrodes, where crack evolution, electric-field redistribution, and resistance signatures are studied together.

%\paragraph{Main contributions.}
The contributions of the paper are as follows:
\begin{enumerate}[leftmargin=*,itemsep=0.3em]
    \item A physically consistent staggered multiphysics formulation combining small-strain linear elasticity, fourth-order phase-field fracture, tensile/compressive energy split, and history-field irreversibility.
    \item A one-way electrical sensing model in which conductivity depends on both strain and damage, allowing gradual piezoresistive drift and abrupt crack-induced resistance change to be represented within a single framework.
    \item A DEM discretization with admissible trial spaces for exact enforcement of essential boundary conditions and quadrature-based evaluation of the variational functionals.
    \item A lean benchmark strategy that validates the electrical building blocks and the fracture engine separately, followed by a coupled application study that constitutes the scientific centerpiece of the paper.
\end{enumerate}

%\paragraph{Paper organization.}
Section~\ref{sec:governing} introduces the governing variational model and the staggered mechanics--fracture--sensing coupling. Section~\ref{sec:dem} presents the DEM discretization, admissible trial functions, quadrature strategy, and implementation remarks. Section~\ref{sec:verification} contains the verification of the building blocks: constant conductivity conduction (E1), uniform-strain piezoresistivity (E2), and a reference-aligned single-edge-notched tension benchmark (M2). Section~\ref{sec:numerical} presents the main numerical study of a tensile plate with stress concentrators and electrodes. Section~\ref{sec:discussion} discusses interpretation, limitations, and outlook, and Section~\ref{sec:conclusion} concludes the paper.

% ============================================================
\section{Governing variational model and staggered multiphysics coupling}
\label{sec:governing}
% ============================================================

\subsection{Domain and primary fields}
Let $\Omega\subset\R^2$ denote the reference configuration of a planar body with boundary $\partial\Omega$. The primary unknowns are the displacement field $\bu:\Omega\to\R^2$, the phase-field variable $\phi:\Omega\to[0,1]$, and the electric potential $V:\Omega\to\R$. The displacement field describes the mechanical response, the phase field regularizes the crack topology through a diffuse representation of fracture, and the electric potential is used to evaluate the self-sensing response once the mechanical and fracture state has been determined.

The mechanical boundary is decomposed into $\Gamma_u\cup\Gamma_t$, where $\bar{\bu}$ is prescribed on $\Gamma_u$ and $\bar{\bt}$ acts on $\Gamma_t$. Likewise, the electrical boundary is decomposed into $\Gamma_V\cup\Gamma_q$, where $\bar V$ is prescribed on $\Gamma_V$ and the current flux $\bar q$ is prescribed on $\Gamma_q$. Insulating boundaries correspond to $\bar q=0$. In the present work, the mechanical and electrical problems are coupled in a staggered one-way sense: fracture evolves under mechanical loading, while the electrical problem is solved afterward on the converged strain and damage fields.

\subsection{Small-strain elasticity}
Under infinitesimal kinematics, the strain tensor is
\begin{equation}
\beps(\bu)=\frac{1}{2}\left(\grad\bu+(\grad\bu)^{\mathsf T}\right).
\end{equation}
For isotropic linear elasticity in plane strain, the Lam\'e constants are
\begin{equation}
\lambda = \frac{E\nu}{(1+\nu)(1-2\nu)},
\qquad
\mu = \frac{E}{2(1+\nu)},
\end{equation}
and the elastic strain-energy density is
\begin{equation}
\psi_e(\beps)=\frac{\lambda}{2}(\tr\beps)^2+\mu\,\beps:\beps.
\end{equation}
This standard choice is sufficient for the present verification benchmarks and for the tensile plate example studied in Section~5. Plane strain is adopted here to remain consistent with the reference-aligned M2 benchmark and to represent a constrained slice of a thicker sensing layer; extending the formulation to plane stress is straightforward but not pursued in the present study. The key nonlinearity in the model does not arise from constitutive plasticity, but from the degradation induced by the evolving fracture field.

\subsection{Tension/compression split and degradation}
To suppress damage growth under compression, we adopt the standard spectral split of the elastic energy. Let $\varepsilon_a$ and $\bn_a$ denote the principal strains and principal directions, and define the ramp operators $\avg{x}_\pm=(x\pm |x|)/2$. Then
\begin{equation}
\beps^\pm = \sum_{a=1}^2 \avg{\varepsilon_a}_\pm \, \bn_a\otimes \bn_a,
\end{equation}
with
\begin{equation}
\psi_e^\pm(\beps)=\frac{\lambda}{2}\avg{\tr\beps}_\pm^2 + \mu\,\beps^\pm:\beps^\pm,
\qquad
\psi_e = \psi_e^+ + \psi_e^-.
\end{equation}
Only the tensile contribution is degraded, so that crack growth is driven by tensile loading while compressive stiffness is retained. The degradation function is chosen as
\begin{equation}
g(\phi)=(1-\phi)^2+\kappa_m,
\qquad 0<\kappa_m\ll 1,
\end{equation}
where the small residual stiffness $\kappa_m$ avoids singular loss of ellipticity in the fully damaged limit. The corresponding degraded mechanical energy density is
\begin{equation}
\psi_m(\bu,\phi)=g(\phi)\,\psi_e^+(\beps(\bu)) + \psi_e^-(\beps(\bu)).
\label{eq:psi_mech}
\end{equation}
Thus, $\phi=0$ represents the intact state, while increasing values of $\phi$ reduce the tensile load-carrying capacity and progressively localize the fracture process.

\subsection{Fourth-order phase-field fracture regularization}
Following the fourth-order AT2-type regularization used in \citep{Goswami2020}, the crack-density contribution is written as
\begin{equation}
\psi_f(\phi)=G_c\left( \frac{\phi^2}{2l} + \frac{l}{4}|\grad\phi|^2 + \frac{l^3}{32}(\Delta \phi)^2 \right).
\label{eq:psi_fracture_density}
\end{equation}
Here, $G_c$ is the critical energy-release rate and $l$ is the regularization length scale controlling the width of the diffused crack zone. Compared with the second-order AT2 form, the fourth-order regularization introduces an additional smoothness contribution through the Laplacian term. This leads to a smoother crack field and is particularly attractive in the DEM setting, where spatial derivatives are obtained directly by automatic differentiation. In the present paper, the fourth-order form is retained to stay aligned with the fracture benchmark that motivates the verification strategy, while the coupled sensing formulation itself remains one-way and physically consistent. For the present sensing setting, the additional smoothness of the fourth-order regularization is also advantageous because it reduces spurious roughness in the damage field and therefore yields more stable conductivity degradation and current-field readout near evolving crack bands.

\subsection{History-field irreversibility}
Fracture irreversibility is enforced through the tensile-energy history field
\begin{equation}
H_n(\bx)=\max_{m\le n}\, \psi_e^+\big(\beps(\bu_m(\bx))\big),
\end{equation}
which stores the maximum previously attained tensile elastic energy at each material point. Within a given load step, $H_n$ is treated as frozen, and after convergence it is updated to the new value of the driving field. This construction prevents crack healing during unloading and avoids the need for explicit inequality constraints on $\phi$. In the present staggered implementation, the history field therefore acts as the crack-driving internal variable that transfers the tensile mechanical state into the fracture subproblem.

\subsection{Mechanical and fracture functionals}
Because the present formulation is solved in staggered form, it is convenient to distinguish between the mechanical equilibrium subproblem and the fracture update subproblem. For a fixed phase field $\phi_n$, the mechanical potential at pseudo-time step $n+1$ reads
\begin{equation}
\Pi_u(\bu\,|\,\phi_n)
=
\int_\Omega \Big[g(\phi_n)\,\psi_e^+(\beps(\bu))+\psi_e^-(\beps(\bu))\Big]\,\mathrm d\Omega
-
\int_{\Gamma_t}\bar{\bt}\cdot\bu\,\mathrm d\Gamma .
\label{eq:Pi_u}
\end{equation}
After the displacement update has been obtained, the fracture subproblem is written in terms of the history field as
\begin{equation}
\Pi_\phi(\phi\,|\,H_n)
=
\int_\Omega \left[
g(\phi)\,H_n
+
G_c\left(
\frac{\phi^2}{2l}
+\frac{l}{4}|\grad\phi|^2
+\frac{l^3}{32}(\Delta\phi)^2
\right)
\right]\mathrm d\Omega .
\label{eq:Pi_phi}
\end{equation}
This staggered decomposition is fully consistent with the variational setting and is also convenient for DEM training: the displacement and phase-field networks can be optimized in alternating fashion while the history field enforces irreversibility across load steps. For later reference, the corresponding total mechanics--fracture contribution may be interpreted as the combination of \eqref{eq:Pi_u} and \eqref{eq:Pi_phi}, rather than as a single monolithic stationarity problem.

\subsection{Electrical self-sensing model}
The electrical problem is treated as a one-way coupled sensing subproblem posed on the converged strain and phase fields. Accordingly, the electric field and current density are
\begin{equation}
\bE = -\grad V,
\qquad
\bJ = \bm{\sigma}(\phi,\beps)\,\bE.
\end{equation}
Under steady conduction, charge conservation requires
\begin{equation}
\grad\cdot \bJ = 0 \qquad \text{in } \Omega.
\end{equation}
The electrical constitutive response is modeled through a resistivity-based piezoresistive update followed by a crack-induced conductivity degradation. We first introduce the in-plane relative resistivity increments
\begin{equation}
\frac{\Delta \rho_{11}}{\rho_0}=\lambda_{11}\varepsilon_{11}+\lambda_{12}\varepsilon_{22},
\qquad
\frac{\Delta \rho_{22}}{\rho_0}=\lambda_{12}\varepsilon_{11}+\lambda_{11}\varepsilon_{22},
\label{eq:delta_rho}
\end{equation}
where $\rho_0=1/\sigma_0$ is the reference resistivity, and $\lambda_{11}$ and $\lambda_{12}$ are the longitudinal and transverse piezoresistive coefficients. The corresponding conductivity components are obtained by inversion,
\begin{equation}
\sigma_{11}(\beps)=\frac{\sigma_0}{1+\Delta\rho_{11}/\rho_0+\eta_r},
\qquad
\sigma_{22}(\beps)=\frac{\sigma_0}{1+\Delta\rho_{22}/\rho_0+\eta_r},
\label{eq:sigma_pz}
\end{equation}
where $\eta_r\ll1$ is a small regularization parameter.

Crack-induced loss of conductivity is then described by the smooth degradation function
\begin{equation}
h_e(\phi;\eta_e,n)=
\frac{1-\exp\!\left[-\eta_e(1-\phi)^n\right]}
     {1-\exp(-\eta_e)+\eta_r}
+\eta_r,
\label{eq:hephi}
\end{equation}
with $\eta_e>0$ controlling the sharpness of the conductivity decay and $n\ge1$ the degradation exponent. This choice satisfies $h_e(0)\approx1$ in the intact state and $h_e(1)\approx0$ in the fully damaged state. The effective in-plane conductivity tensor is therefore taken as
\begin{equation}
\bm{\sigma}(\phi,\beps)=h_e(\phi)\,\mathrm{diag}\!\bigl(\sigma_{11}(\beps),\sigma_{22}(\beps)\bigr).
\label{eq:sigma_eff}
\end{equation}

The electrical field is then obtained from the minimization of the electrical dissipation functional
\begin{equation}
\Pi_{\mathrm e}(V\,|\,\phi,\bu)=\frac{1}{2}\int_\Omega \grad V\cdot \bm{\sigma}(\phi,\beps(\bu))\,\grad V\,\mathrm d\Omega
-\int_{\Gamma_q} \bar q\, V\, \mathrm d\Gamma,
\label{eq:Pi_e}
\end{equation}
subject to the Dirichlet conditions on $\Gamma_V$. Importantly, this electrical problem is solved only after the mechanics--fracture state has converged. The conduction problem therefore does not artificially contribute to crack driving, but acts purely as a sensing/readout mechanism.

\subsection{Resistance extraction}
Once the potential field has been obtained, the signed total current through a prescribed electrode boundary $\Gamma_I\subseteq \Gamma_V$ is computed as
\begin{equation}
I_\Gamma = \int_{\Gamma_I} \bJ\cdot\bn\,\mathrm d\Gamma.
\end{equation}
For resistance reporting, we use the current magnitude,
\begin{equation}
R = \frac{\Delta V}{|I_\Gamma|},
\qquad
\frac{R}{R_0} = \frac{|I_0|}{|I_\Gamma|},
\end{equation}
where $R_0$ and $I_0$ denote the resistance and current of the undamaged reference configuration. In the present setting, the normalized resistance $R/R_0$ is the main sensing observable used throughout the verification and application sections. It provides a compact measure of how strain redistribution and crack evolution jointly modify the global electrical response.

\subsection{Staggered multiphysics solution strategy}
The physically consistent coupling adopted in this work can now be summarized as follows:
\begin{enumerate}[leftmargin=*,itemsep=0.25em]
    \item For a prescribed load increment, solve the mechanics and phase-field subproblems in staggered form until convergence, using the history field to enforce irreversibility.
    \item Update the history field after convergence of the mechanics--fracture step.
    \item Solve the electrical sensing problem on the converged strain and damage fields.
    \item Post-process current, resistance, crack patterns, and energy quantities.
\end{enumerate}
This solution strategy deliberately avoids assigning the electrical conduction problem an artificial crack-driving role. It also mirrors the intended interpretation of the method: fracture is generated by the mechanical state, whereas the electrical field provides an observable signature of the evolving structural condition.

% ============================================================
\section{Deep Energy Method discretization and implementation remarks}
\label{sec:dem}
% ============================================================

\subsection{Admissible neural trial spaces}
In the Deep Energy Method (DEM), the unknown fields are approximated by neural trial functions and the governing variational problem is solved through energy minimization rather than through a conventional finite-element assembly. A key advantage of this setting is that essential boundary conditions can be built directly into the trial space, so that they are satisfied exactly rather than enforced weakly or through penalty terms.

Let $\hat\bu_\theta$, $\hat\phi_\eta$, and $\hat V_\zeta$ denote unconstrained neural-network outputs with trainable parameters $\theta$, $\eta$, and $\zeta$. The admissible displacement and voltage fields are written as
\begin{equation}
\bu(\bx)=\bu_D(\bx)+\mathcal B_u(\bx)\,\hat\bu_\theta(\bx),
\qquad
V(\bx)=V_D(\bx)+\mathcal B_V(\bx)\,\hat V_\zeta(\bx),
\end{equation}
where $\bu_D$ and $V_D$ satisfy the prescribed Dirichlet data and the boundary functions $\mathcal B_u$ and $\mathcal B_V$ vanish on the corresponding Dirichlet boundaries. This construction ensures exact satisfaction of the imposed displacement and potential conditions throughout training.

The phase field is represented by a neural trial function as well. In general, it may be written directly as $\phi(\bx)=\hat\phi_\eta(\bx)$ or through a bounded transformation when this is convenient numerically. For benchmarks with a prescribed starter notch, an additional benchmark-specific seeding or initialization strategy may be introduced to encode the initial crack configuration. In the present paper, this point is relevant mainly for the fracture benchmark M2, where the crack representation must remain consistent with the chosen verification setup.

\subsection{Quadrature-based discrete functionals}
The continuous variational functionals introduced in Section~\ref{sec:governing} are evaluated numerically by composite Gauss quadrature over a partition of the domain into rectangular cells. For a generic integrand $f$, the domain integral is approximated as
\begin{equation}
\int_\Omega f(\bx)\,\mathrm d\Omega
\approx
\sum_{K}\sum_{q\in K} w_q\, f(\bx_q),
\end{equation}
where $\bx_q$ and $w_q$ denote the quadrature points and weights associated with cell $K$. In the DEM setting, these quadrature points play the role of the training locations at which the trial fields and their derivatives are evaluated.

Spatial derivatives required for the strain tensor, the phase-field gradients, and the fourth-order Laplacian term are obtained by automatic differentiation. As a result, the discrete mechanics, fracture, and electrical functionals can be assembled directly from the neural trial fields without introducing element-level interpolation functions in the classical finite-element sense. For the electrical verification benchmarks E1 and E2, fixed composite Gauss grids are sufficient. For the M2 fracture benchmark, the present paper uses a reference-aligned loaded quadrature set associated with the published SENT example of \citep{Goswami2020}. Operationally, this means that the quadrature points and weights for M2 are loaded from the published reference dataset and reused directly, rather than being regenerated within the present manuscript by a new adaptive refinement procedure.

\subsection{Staggered optimization}
At each load step, the mechanics--fracture problem is solved in staggered form, followed by the electrical sensing solve. More specifically, the displacement field is obtained first by minimizing the mechanical subproblem \eqref{eq:Pi_u} for fixed phase field, after which the phase field is updated by minimizing the fracture subproblem \eqref{eq:Pi_phi} for fixed displacement and frozen history field. These two minimizations are alternated until both the relative change in the mechanics--fracture functional and the relative $L^2$ change of the phase field fall below a prescribed tolerance of $10^{-3}$, or until the case-specific maximum number of alternations is reached. The electrical problem is then solved by minimizing the sensing functional \eqref{eq:Pi_e} on the converged strain and damage fields.

This splitting is not merely an implementation convenience. It reflects the intended physics of the present formulation: fracture is driven by the mechanical state, while the electrical field acts as a passive sensing/readout mechanism and therefore should not contribute artificially to crack driving. The staggered structure also aligns naturally with DEM training, because each subproblem can be optimized with its own neural trial space and its own optimizer history.

\subsection{Implementation remarks}
The present manuscript does not claim a new refinement strategy, nor does it present adaptive remeshing as a contribution of its own. Instead, the emphasis is placed on a physically consistent staggered multiphysics formulation and on a compact but meaningful benchmark hierarchy. Accordingly, the electrical benchmarks E1 and E2 are evaluated on fixed composite Gauss grids, whereas the fracture benchmark M2 is performed in a reference-aligned manner using the direct loaded-quadrature route associated with the published SENT example of \citep{Goswami2020}.

In practical terms, the optimization at each load step is performed by a hybrid strategy consisting of an initial Adam stage followed by L-BFGS. Adam provides a robust first descent phase for the neural parameters, while L-BFGS is used to improve final convergence of the variational objective. Warm-starting from the converged solution of the previous load step is employed throughout, which is particularly important for the sequential evolution of the fracture field and for the subsequent electrical sensing solve.

The overall computational sequence used in the present work is summarized in Algorithm~\ref{alg:DEM_summary}.

\subsection{Algorithmic summary}
\label{sec:algorithmic_summary}

Algorithm~\ref{alg:DEM_summary} summarizes the numerical workflow used throughout the paper. Starting from initialized network parameters and an initial history field, the method proceeds load step by load step. At each step, the admissible trial fields are constructed so that essential boundary conditions are satisfied exactly. The variational functionals are then evaluated by quadrature and minimized by the hybrid Adam/L-BFGS strategy. After convergence of the mechanics--fracture stage, the history field is updated and the electrical problem is solved on the converged strain and damage fields. The resulting fields and derived outputs are then stored for post-processing. This summary clarifies the common computational backbone underlying the verification benchmarks of Section~4 and the coupled numerical study of Section~5.

\begin{algorithm}[H]
\caption{DEM workflow for fourth-order phase-field fracture with electrical self-sensing}
\label{alg:DEM_summary}
\begin{algorithmic}[1]
\Require Material and electrical parameters, boundary data, quadrature set
\Ensure Fields $(\mathbf u,\phi,V)$ and derived outputs for all load steps
\State Initialize neural-network parameters and history field
\For{$n=1,\dots,N_s$}
    \State Set the prescribed load/displacement level for step $n$
    \State Warm-start from the converged state of step $n-1$
    \State Construct admissible trial fields with hard Dirichlet enforcement
    \State Evaluate the discrete variational functionals by quadrature
    \State Minimize the mechanics and phase-field subproblems in staggered form using Adam followed by L-BFGS
    \State Check convergence through relative functional and phase-field updates; continue only if the prescribed tolerance is not yet satisfied
    \State Update the history field after convergence
    \State Solve the electrical sensing problem on the converged strain and damage fields
    \State Store fields and derived outputs for post-processing
\EndFor
\end{algorithmic}
\end{algorithm}

% ============================================================
\section{Verification of the building blocks}
\label{sec:verification}
% ============================================================

This section verifies the main ingredients of the proposed formulation with a deliberately compact benchmark set. The electrical subproblem is validated analytically by the constant-conductivity benchmark E1 and the uniform-strain piezoresistive benchmark E2. The fracture component is then verified separately by the single-edge notched tension benchmark M2. This separation keeps the role of each benchmark clear and avoids duplicating the coupled application study presented in Section~5.

\subsection{Electrical verification without fracture}
\label{sec:verification_electrical}

To verify the electrical part of the formulation independently of fracture, we consider two benchmarks in which the phase field is fixed to the intact state, $\phi \equiv 0$. In this regime, the problem reduces to steady conduction with a prescribed conductivity field. Benchmark E1 verifies the electrical variational form, exact enforcement of the electrode boundary conditions, and resistance extraction for constant conductivity. Benchmark E2 verifies the piezoresistive conductivity law under a spatially homogeneous strain state, for which the electrical solution remains analytical.

%\paragraph{Computational setup.}
For both benchmarks, the electrical problem is solved by minimizing the electrical energy functional with exact top--bottom Dirichlet conditions embedded in the trial space. The admissible voltage field is represented by a fully connected neural network with architecture $(2,64,64,64,1)$ and \texttt{tanh} activations. The domain is integrated by composite tensor-product Gauss quadrature using $32\times 32$ cells and three Gauss points per coordinate direction in each cell. Optimization is performed by Adam ($2400$ iterations, learning rate $5\times10^{-3}$), followed by L-BFGS with at most $200$ iterations. Field plots are evaluated on a dense $1001\times1001$ grid, while centerline comparisons use $801$ evaluation points. In Benchmark E2, the conductivity sweep is carried out over $41$ prescribed strain levels in $\varepsilon_{22}\in[0,0.01]$; because the conductivity remains spatially uniform and diagonal, the exact voltage field is unchanged and the same learned admissible potential can be reused throughout the sweep.

\subsubsection{E1: Constant-conductivity conduction}
\label{sec:E1}

Benchmark E1 considers a rectangular domain $\Omega=[0,L]\times[0,W]$ with constant isotropic conductivity $\sigma=\sigma_0$ and top--bottom electrodes. The electric potential is prescribed as
\begin{equation}
V = 0 \quad \text{on } y=0,
\qquad
V = V_{\mathrm{app}} \quad \text{on } y=W,
\qquad
\mathbf{J}\cdot\mathbf{n}=0 \quad \text{on } x=0 \text{ and } x=L,
\label{eq:E1_bc}
\end{equation}
with current density
\begin{equation}
\mathbf{J}=-\sigma_0 \nabla V.
\end{equation}
Because the conductivity is constant, the exact solution is one-dimensional:
\begin{equation}
V(y)=V_{\mathrm{app}}\frac{y}{W},
\qquad
\mathbf{J}=(0,J_y),
\qquad
J_y=-\sigma_0 \frac{V_{\mathrm{app}}}{W}.
\label{eq:E1_exact}
\end{equation}
The corresponding electrode current magnitude per unit thickness is
\begin{equation}
|I|=\sigma_0 \frac{V_{\mathrm{app}}L}{W},
\end{equation}
and the effective resistance is
\begin{equation}
R=\frac{V_{\mathrm{app}}}{|I|}=\frac{W}{\sigma_0 L}.
\label{eq:E1_R}
\end{equation}

\begin{figure}[!htbp]
  \centering
  \includegraphics[width=0.49\linewidth]{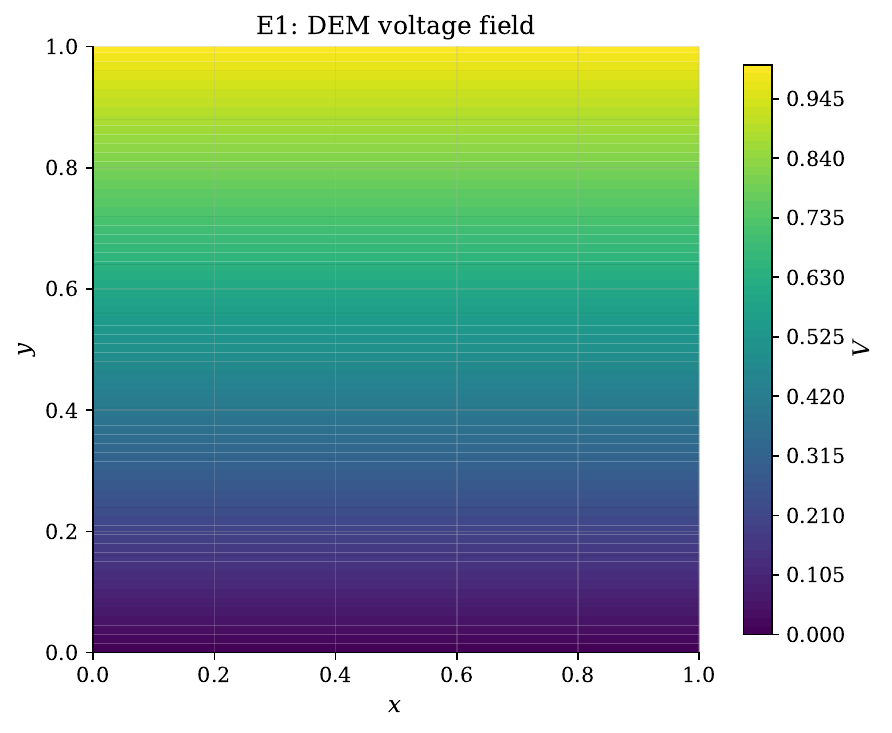}\hfill
  \includegraphics[width=0.45\linewidth]{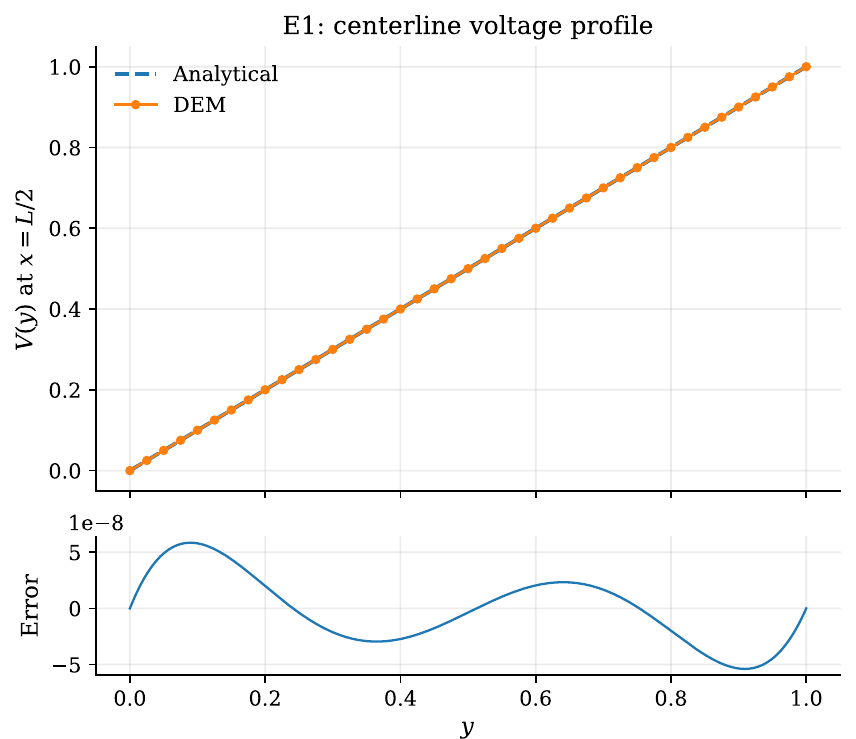}
  \caption{Benchmark E1: constant-conductivity conduction. Left: predicted voltage field $V(x,y)$ for $\sigma=\sigma_0$, showing the expected linear variation in the vertical direction. Right: centerline voltage profile at $x=L/2$ compared with the analytical solution $V(y)=V_{\mathrm{app}}y/W$; the lower panel reports the pointwise error.}
  \label{fig:E1_voltage}
\end{figure}

\begin{figure}[!htbp]
  \centering
  \includegraphics[width=0.47\linewidth]{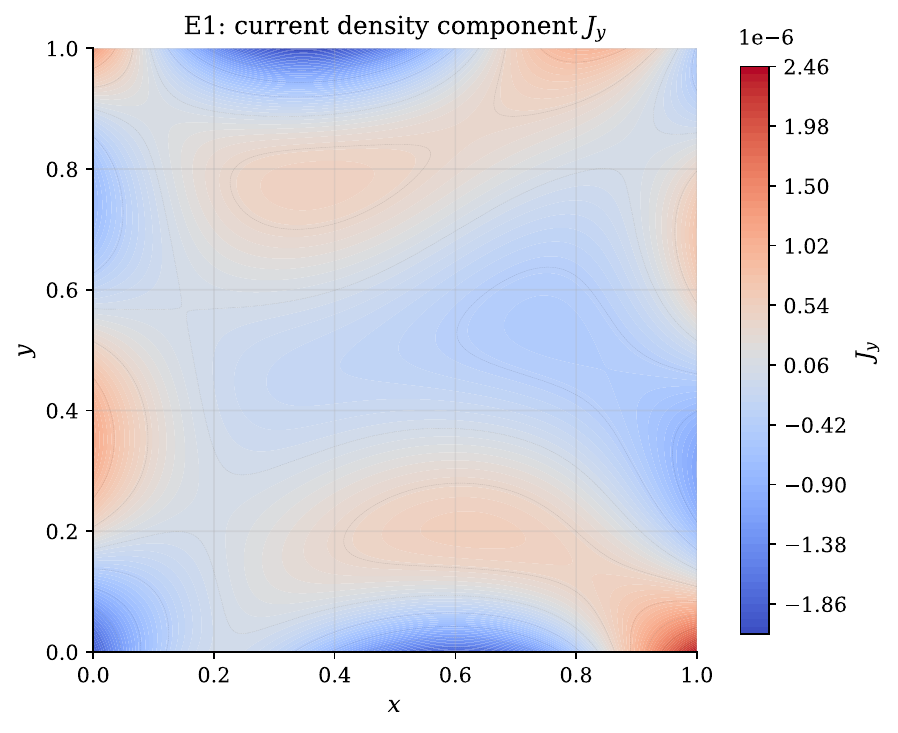}\hfill
  \includegraphics[width=0.52\linewidth]{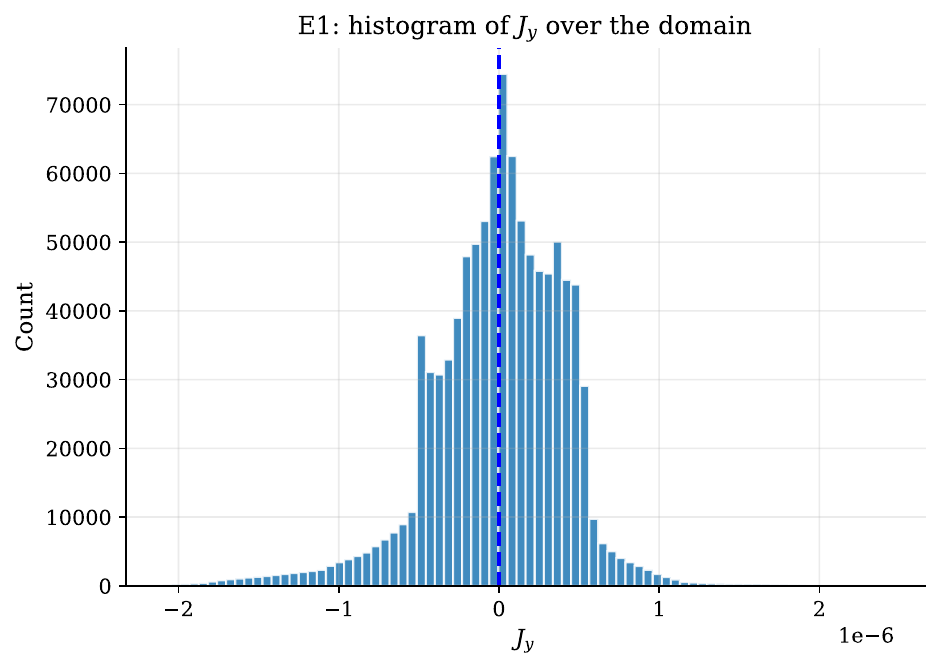}
  \caption{Benchmark E1: current-density verification. Left: numerical deviation of the vertical current-density component from the analytical value, $J_y-J_y^{\mathrm{exact}}$, over the domain. Right: histogram of $J_y-J_y^{\mathrm{exact}}$ over the evaluation grid, showing that the remaining error is concentrated very close to zero.}
  \label{fig:E1_current}
\end{figure}

For the verification run, we choose $L=W=1$, $\sigma_0=1$, and $V_{\mathrm{app}}=1$. The DEM solution reproduces the expected linear voltage field and spatially uniform current density. Figure~\ref{fig:E1_voltage} shows that the predicted potential varies linearly in the vertical direction, while the centerline voltage profile is visually indistinguishable from the analytical solution. Figure~\ref{fig:E1_current} shows the numerical deviation of the vertical current-density component from its analytical constant value, confirming that the remaining current-density error is negligible over the domain. Quantitatively, the relative $L^2$ error of the centerline voltage profile is $5.31\times10^{-8}$. The extracted current is $I_{\mathrm{num}}=-1.00000045$, which gives $R_{\mathrm{num}}=0.99999955$ compared with $R_{\mathrm{exact}}=1.0$ (relative error $4.50\times10^{-7}$). The coefficient of variation of $J_y$ over the evaluation grid is $4.22\times10^{-7}$, and the RMS strong-form residual $\mathrm{RMS}(\nabla\!\cdot\!\mathbf{J})$ is $7.70\times10^{-6}$. These metrics are collected in Table~\ref{tab:E1E2_metrics}. Together, the field plots, profile comparison, and scalar error measures verify the steady-conduction subproblem, the exact enforcement of the electrode boundary conditions, and the current/resistance post-processing.

\FloatBarrier

\subsubsection{E2: Uniform-strain piezoresistivity}
\label{sec:E2}

Benchmark E2 verifies the piezoresistive conductivity law under a spatially homogeneous strain field while retaining the same top--bottom electrode configuration as in E1. Since the strain field is uniform, the conductivity is spatially constant and the electrical solution remains analytical.

We prescribe a uniform vertical strain $\varepsilon_{22}$ together with the corresponding Poisson contraction
\begin{equation}
\varepsilon_{11}=-\nu \varepsilon_{22},
\end{equation}
so that $\varepsilon$ is constant throughout $\Omega$ and $\phi\equiv 0$. Under the boundary conditions in~\eqref{eq:E1_bc}, the exact potential remains
\begin{equation}
V(y)=V_{\mathrm{app}}\frac{y}{W}.
\end{equation}
Using the resistivity-based piezoresistive update introduced in Section~2, and noting that $\phi\equiv0$ so that $h_e(\phi)=1$, the conductivity component relevant for top--bottom conduction is
\begin{equation}
\sigma_{22}(\varepsilon)=\frac{\sigma_0}{1+\Delta\rho_{22}/\rho_0+\eta_r},
\qquad
\frac{\Delta\rho_{22}}{\rho_0}=\lambda_{12}\varepsilon_{11}+\lambda_{11}\varepsilon_{22}.
\label{eq:E2_sigma}
\end{equation}
Hence the closed-form effective resistance is
\begin{equation}
R(\varepsilon_{22})=\frac{W}{\sigma_{22}(\varepsilon)L},
\end{equation}
and the normalized resistance scales as
\begin{equation}
\frac{R(\varepsilon_{22})}{R_0}
=
\frac{\sigma_{22}(0)}{\sigma_{22}(\varepsilon)}.
\label{eq:E2_R}
\end{equation}

\begin{figure}[!htbp]
  \centering
  \includegraphics[width=0.6\linewidth]{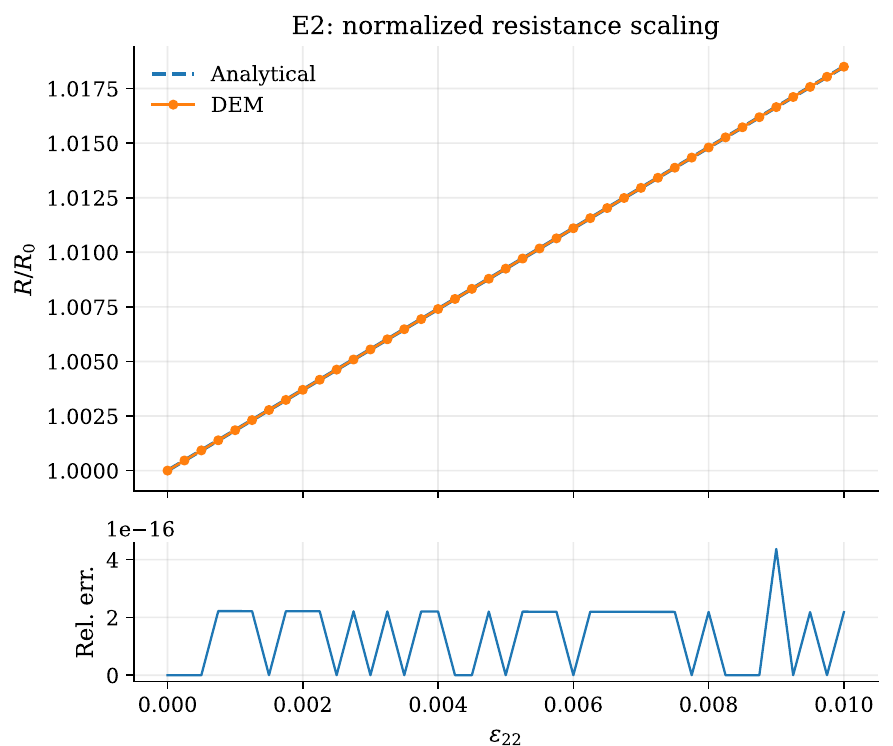}
  \caption{Benchmark E2: uniform-strain piezoresistivity with top--bottom electrodes. Normalized resistance $R/R_0$ versus prescribed uniform strain $\varepsilon_{22}$. The DEM prediction is visually indistinguishable from the analytical scaling; the lower panel reports the relative error.}
  \label{fig:E2_R_scaling}
\end{figure}

\begin{figure}[t]
  \centering
  \includegraphics[width=0.58\linewidth]{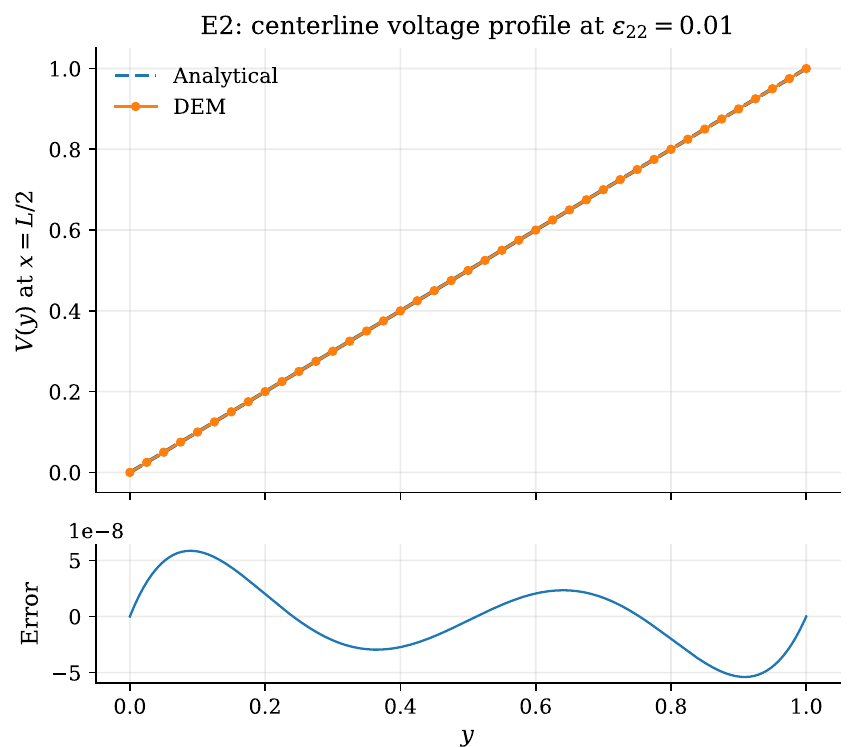}
  \caption{Benchmark E2: centerline voltage profile at the maximum prescribed strain, $\varepsilon_{22}=0.01$. The potential remains linear in $y$, as expected for spatially uniform conductivity; the lower panel shows the pointwise error.}
  \label{fig:E2_voltage_profile}
\end{figure}

For the verification sweep, we use $\varepsilon_{22}\in[0,0.01]$, $\nu=0.3$, $\lambda_{11}=2.0$, and $\lambda_{12}=0.5$. Figure~\ref{fig:E2_R_scaling} shows that the normalized resistance predicted by the DEM solver is visually indistinguishable from the analytical scaling over the full strain range; the lower panel confirms that the remaining discrepancy is at machine precision. Figure~\ref{fig:E2_voltage_profile} shows that the centerline voltage profile at the maximum prescribed strain remains linear in $y$, as expected for spatially uniform conductivity. Quantitatively, the DEM prediction preserves the linear voltage profile throughout the strain sweep, with mean centerline-profile error $5.31\times10^{-8}$ and mean coefficient of variation $\mathrm{CV}(J_y)=4.22\times10^{-7}$. Most importantly, the normalized resistance response follows the analytical scaling in~\eqref{eq:E2_R} essentially exactly, with a maximum relative deviation of $4.37\times10^{-16}$ over the full strain range. These values are summarized in Table~\ref{tab:E1E2_metrics}. This benchmark therefore verifies the piezoresistive conductivity update and confirms that the resistance extraction remains consistent when the conductivity changes through strain alone.

\begin{table}[!htbp]
\centering
\caption{Electrical verification metrics for benchmarks E1 and E2.}
\label{tab:E1E2_metrics}
\begin{tabular}{llc}
\toprule
Benchmark & Quantity & Value \\
\midrule
\multicolumn{3}{l}{\textit{Voltage field}}\\
E1 & relative $L^2$ error of centerline voltage & $5.31\times10^{-8}$ \\
E2 & mean relative $L^2$ error of centerline voltage & $5.31\times10^{-8}$ \\
\midrule
\multicolumn{3}{l}{\textit{Current field}}\\
E1 & relative $L^2$ error of $J_y$ & $4.22\times10^{-7}$ \\
E1 & $\mathrm{CV}(J_y)$ & $4.22\times10^{-7}$ \\
E1 & $\mathrm{RMS}(\nabla\!\cdot\!\mathbf J)$ & $7.70\times10^{-6}$ \\
E2 & mean $\mathrm{CV}(J_y)$ & $4.22\times10^{-7}$ \\
\midrule
\multicolumn{3}{l}{\textit{Resistance}}\\
E1 & numerical resistance $R_{\mathrm num}$ & $0.99999955$ \\
E1 & relative resistance error & $4.50\times10^{-7}$ \\
E2 & max. relative error in $R/R_0$ & $4.37\times10^{-16}$ \\
\bottomrule
\end{tabular}
\end{table}

The repeated values appearing for some voltage- and current-related metrics in E1 and E2 are a consequence of this setup: once the admissible potential has been learned for the homogeneous top--bottom conduction problem, the centerline voltage shape remains unchanged throughout the E2 strain sweep.

\FloatBarrier

\subsection{M2: Reference-aligned single-edge-notched tension benchmark}
\label{sec:m2}
The fracture part of the formulation is verified by a single-edge-notched tension (SENT) benchmark aligned with the unit-square fourth-order DEM example of \citep{Goswami2020}. The purpose of M2 in the present paper is deliberately focused. Rather than serving as a broad benchmarking study, it is used to confirm that the fracture solver produces (i) crack initiation from the prescribed notch, (ii) predominantly mode-I propagation along the expected horizontal path, and (iii) a sensible loss of load-carrying capacity before the coupled tensile-plate study of Section~\ref{sec:numerical} is introduced.

The specimen is a unit square plate with an initial horizontal notch extending from the left boundary to mid-width. The kinematic boundary conditions follow the admissible trial functions used in the reference implementation: the horizontal displacement is suppressed on both vertical edges, while the vertical displacement is prescribed at the top and bottom boundaries. To keep the benchmark reproducible, we use the direct loaded-quadrature route together with the published reference quadrature set. A nonuniform displacement schedule is employed so that the solution is sampled more densely around the onset of softening while still containing the reference-style snapshot levels $\bar v=10^{-3}$, $4\times10^{-3}$, and $6\times10^{-3}$. The fracture network uses the reference-aligned architecture $(2,50,50,50,3)$ and is trained at each load step by Adam followed by L-BFGS.

\begin{figure}[!htbp]
  \centering
  \includegraphics[width=0.99\linewidth]{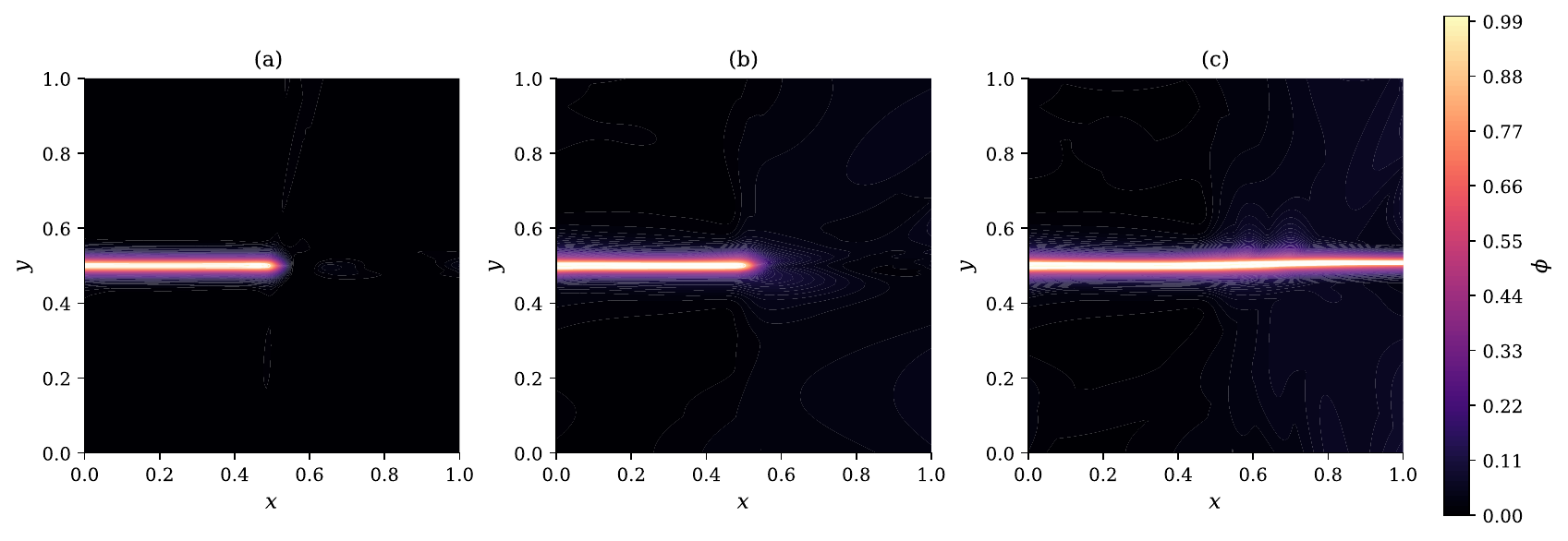}
  \caption{M2 benchmark: selected phase-field snapshots for the reference-aligned SENT example at prescribed displacements $\bar v=10^{-3}$, $4\times10^{-3}$, and $6\times10^{-3}$. The crack initiates from the starter notch and evolves predominantly in the horizontal direction, consistent with the expected mode-I crack path.}
  \label{fig:M2_snapshots}
\end{figure}

Figure~\ref{fig:M2_snapshots} summarizes the crack evolution in a compact three-panel snapshot view. At the earliest level, $\bar v=10^{-3}$, the phase field remains concentrated around the starter notch and only a limited forward extension is visible. This is the expected behaviour at small load, where the crack tip is activated but the damaged region remains close to the initial defect. At $\bar v=4\times10^{-3}$, the crack has advanced markedly beyond the initial notch and remains centered around the horizontal midline of the specimen. This intermediate state is particularly important, because it shows that the solver drives the crack in the correct direction rather than producing spurious diagonal branching or diffuse off-axis damage. At $\bar v=6\times10^{-3}$, the damaged region spreads into a broader horizontal band. The field is therefore no longer a perfectly sharp crack representation, but the dominant evolution still follows the notch axis and remains consistent with the expected mode-I propagation pattern of the benchmark. The snapshot panel is thus interpreted primarily as a \emph{fracture-pattern verification}: the key evidence is notch-driven horizontal growth, not an exact pointwise match of every contour level with the reference article.

%The global response is shown in Figure~\ref{fig:M2_response}. The reaction force increases smoothly from the initial loading stage, reaches a peak value of approximately $5.97\times10^{2}$ at $\bar v\approx 4.4\times10^{-3}$, and then drops rapidly as the fracture process localizes. This rise-and-drop behaviour is the expected qualitative signature of stiffness loss due to crack growth. At the same time, the direct loaded-quadrature route becomes less reliable deep in the far post-peak regime, where the force history begins to fluctuate and eventually becomes weakly nonphysical. For that reason, only the trimmed, physically meaningful portion of the force--displacement response is retained in the manuscript. This trimmed representation is sufficient for the purpose of Section~\ref{sec:verification}: together with the snapshot panel, it shows that the present fracture solver (a) activates at the correct notch location, (b) propagates predominantly along the expected horizontal path, and (c) exhibits a qualitatively correct softening response once the crack has advanced.

\begin{figure}[!htbp]
  \centering
  \includegraphics[width=0.60\linewidth]{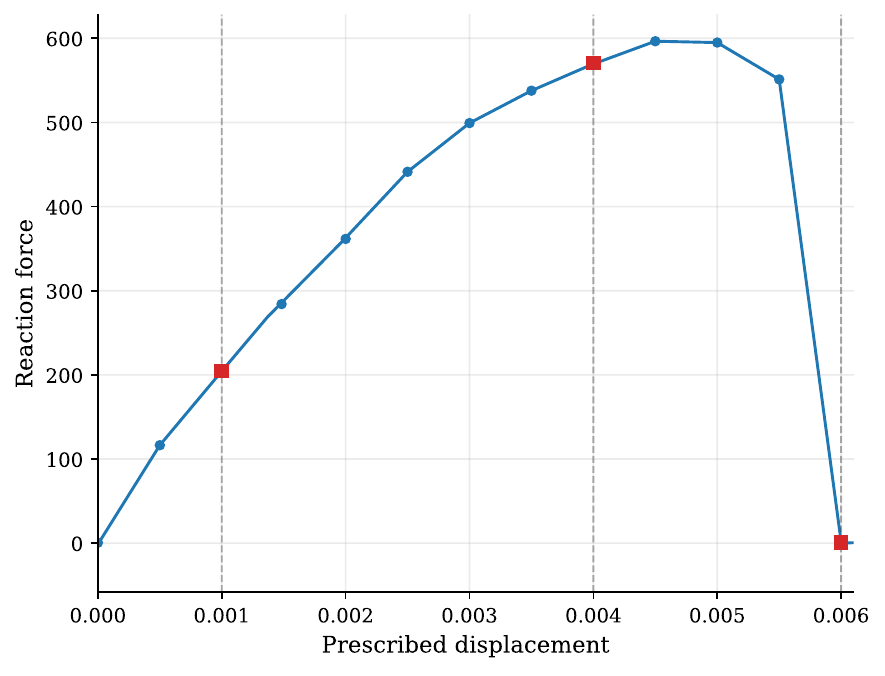}
  \caption{M2 benchmark: force--displacement response for the reference-aligned SENT example. The reaction force increases smoothly to a peak and then drops rapidly as the crack localizes.}
  \label{fig:M2_response}
\end{figure}

The global response is shown in Figure~\ref{fig:M2_response}. The reaction force increases smoothly from the initial loading stage, reaches a peak value of approximately $5.97\times10^{2}$ at $\bar v\approx 4.4\times10^{-3}$, and then drops rapidly as the fracture process localizes. This rise-and-drop behaviour is the expected qualitative signature of stiffness loss due to crack growth. For the present verification purpose, the most relevant part of the response is the regime around crack initiation and early post-peak softening, where the mechanical evolution remains directly linked to the notch-driven fracture process. Together with the snapshot panel, the force--displacement curve shows that the present fracture solver (a) activates at the correct notch location, (b) propagates predominantly along the expected horizontal path, and (c) exhibits a qualitatively correct softening response once the crack has advanced. Table~\ref{tab:M2_compare} supplements this qualitative assessment by comparing the peak force and the displacement at peak with the reference values reported in~\citep{Goswami2020}.

Taken together, Figures~\ref{fig:M2_snapshots} and~\ref{fig:M2_response} verify that the fracture solver captures the essential behavior of the reference-aligned SENT problem: notch activation, predominantly horizontal mode-I propagation, and a clear loss of load-carrying capacity after crack growth becomes pronounced. In this sense, M2 provides the required fracture-engine verification before the electrically instrumented tensile-plate study of Section~\ref{sec:numerical}.

\begin{table}[!htbp]
\centering
\caption{Small quantitative comparison for the M2 SENT benchmark.}
\label{tab:M2_compare}
\begin{tabular}{lccc}
\toprule
Quantity & Present DEM & Reference [19] & Relative deviation \\
\midrule
Peak reaction force & $5.97\times 10^2$ & $6.20\times 10^2$ & $-3.71\%$ \\
Displacement at peak & $4.4\times 10^{-3}$ & $4.2\times 10^{-3}$ & $+4.76\%$ \\
\bottomrule
\end{tabular}
\end{table}

%\paragraph{Role of Section~\ref{sec:verification}.}
The verification section is intentionally compact. E1 and E2 validate the electrical building blocks analytically, while M2 validates the fracture engine against a reference-aligned SENT benchmark. With these three checks in place, the main coupled study can focus on the scientific questions of interest rather than on broad method benchmarking.

\FloatBarrier

% ============================================================
\section{Numerical study: tensile plate with stress concentrators and electrodes}
\label{sec:numerical}
% ============================================================

This section is the scientific centerpiece of the paper. Its purpose is to demonstrate, in a geometry that is richer than the verification cases of Section~\ref{sec:verification}, how evolving fracture patterns interact with electrical fields and how this interaction is reflected in global electrical observables. The example considered is a tensile plate containing multiple stress concentrators and instrumented by boundary electrodes. The goal is not to provide an exhaustive parameter study, but to show in a controlled and reproducible setting how crack growth, current-path redistribution, and resistance change can be interpreted together.

\subsection{Geometry, loading, and electrode configuration}
A square plate of dimensions, $L=W=1~\mathrm{mm}$, is considered under displacement-controlled tension. The lower boundary is fixed in the vertical direction, the upper boundary is prescribed a monotonically increasing vertical displacement, and the left and right boundaries are kinematically restrained in the horizontal direction, consistent with the admissible trial-space construction used in the implementation. Electrically, a potential difference is applied between top and bottom electrodes so that the vertical current path interacts with the evolving damage field.

\begin{figure}[!htbp]
  \centering
  \includegraphics[width=0.4\linewidth]{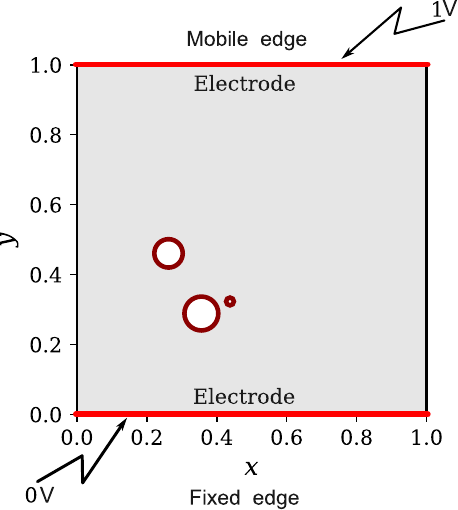}
  \caption{Illustrative geometry and dominant boundary conditions for the tensile plate with stress concentrators and electrodes used in Section~\ref{sec:numerical}. The schematic emphasizes the applied tensile loading and the top--bottom electrode configuration. The horizontal side restraints used in the computation are not drawn explicitly, but are accounted for in the admissible trial-space construction described in Section~\ref{sec:dem}.}
  \label{fig:S5_geometry}
\end{figure}

To create a reproducible but nontrivial fracture pattern, three circular holes are embedded in the plate and act as stress concentrators. In the present study the hole geometry is frozen rather than randomly varied, so that the numerical response is tied to a single, explicitly defined specimen. All geometric coordinates and radii reported below are given in mm. The hole centers and radii are
\[
(c_x,c_y,r)=
(0.262,\,0.460,\,0.0406),\qquad
(0.356,\,0.288,\,0.0485),\qquad
(0.438,\,0.323,\,0.0111).
\]
The specimen together with the dominant loading and electrode arrangement is shown schematically in Figure~\ref{fig:S5_geometry}; the horizontal side restraints used in the computation are omitted from the drawing for visual clarity. The chosen configuration creates competing ligaments and nontrivial current paths, which makes it suitable for examining how local crack growth translates into a global resistance signature.

\subsection{Material parameters and numerical settings}
The material and numerical parameters used in the numerical study are collected in Table~\ref{tab:params}. The mechanical model uses linear elasticity together with the fourth-order phase-field fracture regularization introduced in Section~\ref{sec:governing}. The tensile-plate study uses a fully connected neural network with architecture $(2,50,50,50,4)$, where the outputs correspond to the horizontal displacement, vertical displacement, phase field, and electric potential. The computations are carried out in double precision. Domain integrals are evaluated on a $40\times 40$ cell partition with $4\times 4$ Gauss points per cell, while post-processing fields are evaluated on a $100\times 100$ prediction grid filtered outside the holes. The loading history consists of $N_s=35$ displacement-controlled steps with increment $\Delta\bar v = 8\times10^{-4}$. At each step, the network parameters are warm-started from the converged state of the previous load step and optimized using 500 Adam iterations followed by 500 L-BFGS iterations. A global random seed of 1234 is used for the solver and a fixed seed of 111 is used for the hole layout, so that the reported geometry and response are fully reproducible. In the Section~\ref{sec:numerical} runs, the phase field is not hard-bounded during optimization; mild local overshoots are therefore clipped only for visualization and interpreted as numerical artifacts rather than physical values. In the present context, the resulting electrical contribution is interpreted diagnostically as a numerical indicator of crack--signal interaction rather than as a crack-driving energetic mechanism.

\begin{table}[!htbp]
  \centering
  \caption{Material and numerical parameters for the tensile-plate study.}
  \label{tab:params}
  \begin{tabular}{lll}
    \toprule
    Quantity & Symbol & Value \\
    \midrule
    Young's modulus & $E$ & $5.0\times10^4~\mathrm{MPa}$ \\
    Poisson ratio & $\nu$ & $0.3$ \\
    Fracture toughness & $G_c$ & $1.7~\mathrm{N/mm}$ \\
    Phase-field length scale & $l$ & $0.005~\mathrm{mm}$ \\
    Reference conductivity & $\sigma_0$ & $1.0$ \\
    Piezoresistive coefficients & $\lambda_{11},\lambda_{12}$ & $2.0,\;0.5$ \\
    Electrical degradation sharpness & $\eta_e$ & $50.0$ \\
    Damage-degradation exponent & $n$ & $6.0$ \\
    Electrical regularization & $\eta_r$ & $10^{-8}$ \\
    Applied voltage drop & $V_{\mathrm app}$ & $1.0~\mathrm{V}$ \\
    Number of holes & -- & $3$ circular holes \\
    Network architecture & -- & $(2,50,50,50,4)$ \\
    Arithmetic precision & -- & float64 \\
    Quadrature mesh & -- & $40\times40$ cells \\
    Gauss rule & -- & $4\times4$ points/cell \\
    Prediction grid & -- & $100\times100$ filtered outside holes \\
    Number of load steps & $N_s$ & $35$ \\
    Load increment & $\Delta \bar v$ & $8\times10^{-4}$ \\
    Adam iterations / step & -- & $500$ \\
    L-BFGS iterations / step & -- & $500$ \\
    Warm start & -- & previous converged load step \\
    Computational platform & -- & single-workstation TensorFlow implementation \\
    Global solver seed & -- & $1234$ \\
    Hole-layout seed & -- & $111$ \\
    \bottomrule
  \end{tabular}
\end{table}

\subsection{Selected field evolution and crack--signal interaction}
\begin{figure}[!htbp]
  \centering
  \includegraphics[width=\linewidth]{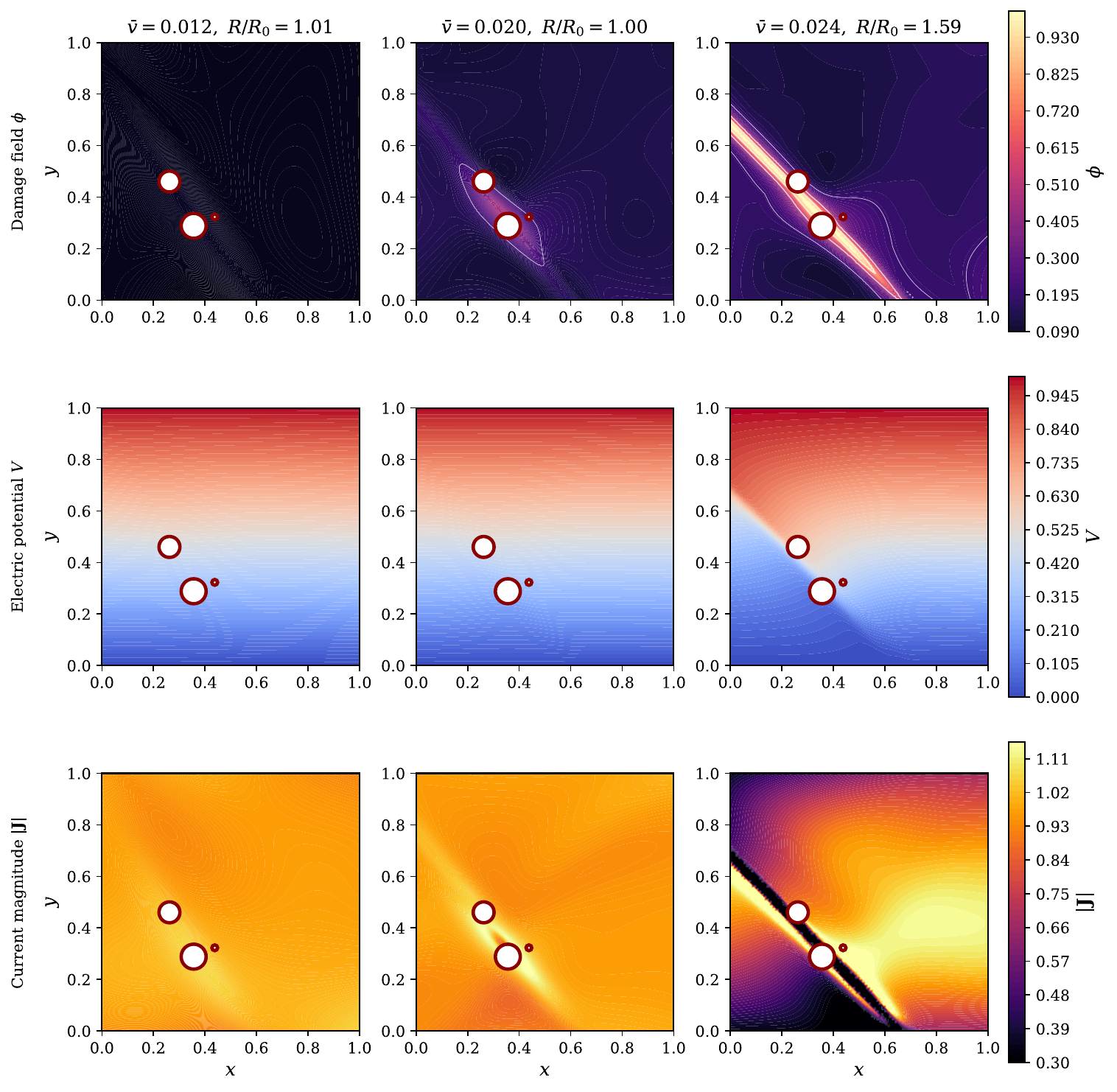}
  \caption{Curated multiphysics field evolution for the tensile plate with stress concentrators. The rows correspond to three selected load levels, $\bar v=0.012$, $0.020$, and $0.024$. The columns show the damage field $\phi$, the electric potential $V$, and the current-density magnitude $|\mathbf J|$. The sequence highlights the transition from weak localization with minor sensing changes to a strongly developed damage channel associated with pronounced current-path disruption.}
  \label{fig:S5_multiphysics}
\end{figure}

Figure~\ref{fig:S5_multiphysics} presents the main field-level result of the paper. Three load levels are shown, corresponding to an early stage with weak localization, an intermediate stage with clearly developed damage around the hole cluster, and a later stage in which a dominant damaged channel strongly perturbs the conducting paths. For each selected state, the figure reports the damage field $\phi$, the electric potential $V$, and the current-density magnitude $|\mathbf J|$.

At the earliest selected state, $\bar v=0.012$, the damage field remains weakly localized, with a maximum value of approximately $\phi_{\max}\approx 0.075$, and the electrical response is only mildly perturbed. The normalized resistance is still close to the initial value, with $R/R_0\approx 1.014$. At this stage, the current can still traverse the specimen through several alternative ligaments, so the electrical signature remains small even though the stress concentrators already influence the mechanical field.

At the intermediate level, $\bar v=0.020$, the damage pattern has expanded significantly and a clearer preferential fracture channel begins to form around the holes. This is accompanied by a visible redistribution of the potential field and a more focused rerouting of the current-density magnitude around the developing damaged zone. Interestingly, the normalized resistance is still close to unity, with $R/R_0\approx 0.997$. This indicates that moderate crack growth does not necessarily produce an immediately large global resistance change if the remaining conductive ligaments are still capable of carrying current efficiently.

At the later selected state, $\bar v=0.024$, the damaged region has become sufficiently strong to disrupt the dominant current paths. The voltage contours become more distorted, the current-density field is forced into narrower remaining ligaments, and the global resistance rises sharply to $R/R_0\approx 1.588$. This stage is the clearest illustration of the central message of the section: the electrical signal is not simply a monotone proxy for local damage magnitude, but reflects the geometrically constrained reorganization and eventual blockage of the current pathways.

\subsection{Global response and interpretation}
\begin{figure}[!htbp]
  \centering
  \includegraphics[width=0.7\linewidth]{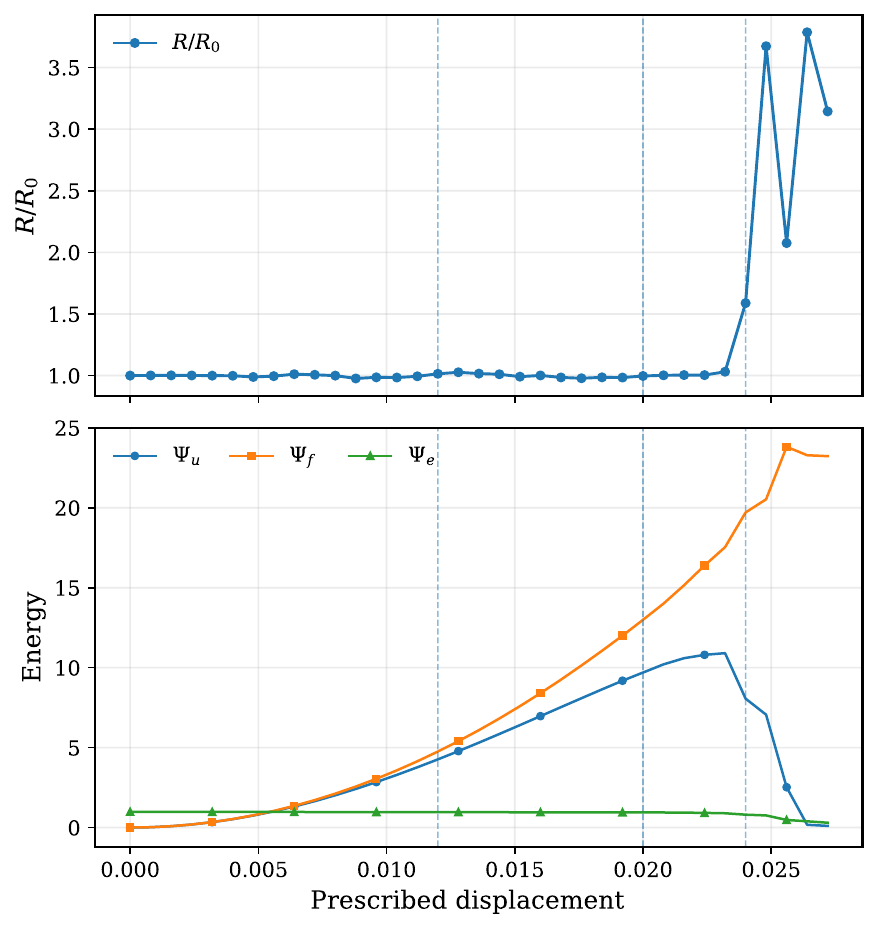}
  \caption{Global response of the tensile plate with stress concentrators. Top: normalized resistance $R/R_0$ versus prescribed displacement. Bottom: evolution of the mechanical, fracture, and electrical energy contributions. The vertical markers identify the three load levels visualized in Figure~\ref{fig:S5_multiphysics}.}
  \label{fig:S5_global}
\end{figure}

The corresponding global response is shown in Figure~\ref{fig:S5_global}. The upper panel reports the normalized resistance $R/R_0$ as a function of the prescribed displacement, while the lower panel shows the evolution of the mechanical, fracture, and electrical energy contributions. In the present context, the electrical contribution is interpreted diagnostically as an indicator of crack--signal interaction, not as evidence that the electrical field drives fracture. The vertical markers indicate the three states selected in Figure~\ref{fig:S5_multiphysics}.

The global curves support a two-regime interpretation of the sensing response. In the early part of the loading history, the resistance varies only mildly, despite the fact that damage has already begun to localize around the stress concentrators. This is consistent with a regime in which strain redistribution and modest conductivity changes remain partially compensated by alternative current paths. Once a dominant damaged channel emerges and begins to sever the most efficient conductive ligaments, the resistance rises much more sharply. The late-stage rise in $R/R_0$ is therefore best interpreted as a current-path disruption effect rather than as a direct pointwise image of the phase field.

The energy curves provide a consistent companion interpretation. The mechanical part evolves smoothly at early stages, while the fracture contribution grows more strongly once localization becomes pronounced. The electrical contribution changes in parallel with the increasing field distortion. The strongest late-stage oscillations in $R/R_0$ are interpreted cautiously: they are consistent with abrupt switching of the remaining conductive ligaments, but they are also likely amplified by reduced numerical robustness once the conducting path becomes highly localized and fragmented. Together, these trends confirm that the global electrical response is most informative when interpreted in conjunction with the fracture evolution rather than in isolation.

\subsection{Summary of the numerical study}
The tensile-plate study demonstrates that the proposed framework can generate an interpretable crack--signal interaction picture in a geometry that is more application-relevant than the benchmark cases of Section~\ref{sec:verification}. The most important finding is not merely that resistance changes when fracture evolves, but that the character of the signal depends on the manner in which conductive paths are redistributed and eventually blocked. In this sense, the numerical study supports a mechanism-based interpretation of resistance-based self-sensing: gradual changes may occur during early localization, whereas stronger global resistance changes emerge once a dominant current-carrying ligament is disrupted.

\FloatBarrier

% ============================================================
\section{Discussion}
\label{sec:discussion}
% ============================================================

The numerical evidence of this paper should be interpreted in light of the modeling choices made. First, the one-way coupling strategy adopted in Sections~\ref{sec:governing}--\ref{sec:verification} is deliberate. For passive self-sensing materials under low electrical excitation, the electrical field is most naturally interpreted as a readout of the evolving mechanical and fracture state rather than as an energetic driving mechanism. This avoids arbitrary weighting between mechanical and electrical energies and keeps the physical interpretation transparent.

Second, the conductivity law used here is intentionally simple. The linearized piezoresistive representation is sufficient for verifying the implementation in E1 and E2 and for revealing meaningful crack--signal interaction trends in Section~\ref{sec:numerical}, but it does not resolve all mesoscale transport mechanisms of conductive networks. In realistic materials, experimental calibration may require richer constitutive laws, anisotropic conductivity evolution, or homogenized models that account explicitly for network topology and percolation effects.

Third, the benchmark section validates the electrical building blocks through E1 and E2 and the fracture engine through a reference-aligned SENT example. This provides a coherent basis for the Section~\ref{sec:numerical} study, where the interaction between stress concentrators, crack development, and electrical-current rerouting becomes visible in a single example and can be interpreted directly through the resulting resistance signal.

Finally, the global resistance signal must not be interpreted as a one-to-one surrogate for local damage. The present numerical study shows that resistance can remain nearly unchanged during stages of appreciable damage development if sufficiently conductive ligaments remain available, and can increase sharply once those ligaments are compromised. This observation is important for future sensing design, inverse identification, and learning-enabled structural health monitoring workflows, because it highlights the need to interpret global electrical observables through the geometry of the evolving current paths.

% ============================================================
\section{Conclusions}
\label{sec:conclusion}
% ============================================================

A physically consistent DEM framework has been presented for fourth-order phase-field fracture with piezoresistive self-sensing. The mechanics--fracture part combines small-strain linear elasticity, tensile/compressive energy split, history-field irreversibility, and a fourth-order AT2-type crack regularization. The electrical part is formulated as a sensing problem in which conductivity depends on strain and damage, thereby providing a global observable that can be compared directly with the evolving fracture state.

The manuscript has been organized around a deliberately lean verification strategy. Two analytical benchmarks validate the electrical building blocks, and one reference-aligned single-edge-notched tension benchmark validates the fracture component. This compact hierarchy provides a sufficient methodological basis for the main application study without turning the paper into a broad solver-comparison exercise.

The numerical study of the perforated tensile plate is the central contribution of the paper. It shows that damage localization around stress concentrators does not automatically produce a large global resistance change; instead, the resistance signal becomes most pronounced once dominant conductive ligaments are disrupted and current paths are forced to reorganize strongly. In this sense, the proposed framework supports a mechanistic interpretation of resistance-based self-sensing in fractured structures.

The broader significance of the paper lies in this connection between structural fracture and electrical observability. The framework can serve not only as a forward simulation tool, but also as a generator of synthetic coupled fracture--sensing datasets for inverse identification, sensor-layout design, and future learning-enabled structural health monitoring workflows.

% ============================================================
%Acknowledgment
% ============================================================
%\section*{Acknowledgment}
%This work stems from the research project “Functionalized, Multi-Physically Optimized Adhesive Systems for Inherent Structural Monitoring of Rotor Blades” (Func2Ad – Funktionalisierte, multiphysikalisch optimierte Klebstoffsysteme für die inhärente Strukturüberwachung von Rotorblättern), funded by the Federal Ministry for Economic Affairs and Climate Action, Germany (Grant No. 03EE3069A). The authors gratefully acknowledge this financial support. They also acknowledge the use of the LUIS scientific computing cluster, Germany, funded by Leibniz Universität Hannover, the Lower Saxony Ministry of Science and Culture (MWK), and the German Research Foundation (DFG).

\section*{Compliance with ethical standards}

\noindent \textbf{Data availability statement:}\\
The data supporting the findings of this study are available from the corresponding author upon reasonable request.
%All scripts, input files, quadrature datasets, and post-processing utilities used for the numerical examples will be archived in a public repository upon acceptance. A persistent DOI-linked archive together with the version-controlled source repository will be provided in the final published version.

\noindent \textbf{Funding statement:}\\
This research did not receive any specific grant from funding agencies in the public, commercial, or not-for-profit sectors.

\noindent \textbf{Conflict of interest disclosure:}\\
The authors declare that they have no known competing financial interests or personal relationships that could have appeared to influence the work reported in this paper.

%\noindent \textbf{Ethics approval statement:}\\
%Not applicable.

%\noindent \textbf{Patient consent statement:}\\
%Not applicable.

%\noindent \textbf{Permission to reproduce material from other sources:}\\
%Not applicable.

%\noindent \textbf{Clinical trial registration:}\\
%Not applicable.

\noindent \textbf{Declaration of AI-assisted technologies in the writing process:}\\
During the preparation of this manuscript, the authors utilized AI-based tools to assist with language refinement and grammar checking. Following the use of these tools, the authors thoroughly reviewed and edited the content, and take full responsibility for the final version of the publication.

% ============================================================
\bibliographystyle{elsarticle-num}

\bibliography{references}

\end{document}